# Spinon Superconductivity and Superconductivities Mediated by Spin-Waves and Phonons in Cuprates.

A. Mourachkine

*Université Libre de Bruxelles, Service de Physique des Solides, CP233, Boulevard du Triomphe, B-1050 Brussels, Belgium*

The disclosure of spinon superconductivity and superconductivity mediated by spin-waves in hole-doped $Bi_2Sr_2CaCu_2O_{8+x}$ (Bi2212) cuprate raises the question about the origin of the superconductivity in other cuprates and specially in an electron-doped $Nd_{2-x}Ce_xCuO_4$ (NCCO) cuprate.

The superconductivity (SC) in copper-oxide materials or the high-$T_c$ superconductivity (HTSC) is an extraordinary phenomenon. The cuprates is the first system with the coexistence of two different types of the SC with the different mechanisms [2]. The most astonishing fact is that the coherent state of the spinon SC in hole-doped cuprates is established via another type of the SC [2] mediated by spin-waves (magnons) [3]. The SC due to pairing of spinons we will call the spinon SC. Spinons are chargeless fermions which occur in one-dimensional physics [4]. The spinon SC occurs on charged stripes into $CuO_2$ planes [5]. The stripes are separated by antifferomagnetic (AF) isolating domains. The spinons create pairs on stripes because of the existence of a large spin-gap due to AF correlations [6]. The $2\Delta/k_BT_c$ value is a very important characteristic of the SC. It indicates the coupling strength of Copper pairs. In conventional superconductors, this value is about 3.5. In Bi2212, the $2\Delta/k_BT_c$ value for spinon SC can be incredibly high, up to 30 [2]. The coherence length of the spinon pairs can be very short since spinons have no problem with the Coloumb repulsive force. At the same time, it is a huge problem for the spinon SC in order to established the coherent-SC state because spinons are located on stripes which are separated. The distance between charged stripes is too large for the spinon SC in order to establish the coherent state. Nature had to look for another mechanism "to connect" the charged stripes. In hole-doped cuprates, it happened that spin-waves in AF regions [3] are the most suitable mechanism for it. Since electron-magnon interactions are much stronger than electron-phonon interactions in AF compounds [3], the $T_c$ can be much higher than in classical superconductors. At $T_c$, the coherent state of the spinon SC into $CuO_2$ planes occurs due to and via the SC mediated by spin-waves, which can be considered as pairing of magnetic polarons [3]. The magnetic polaron (MP) can be pictured as the electron and a spin polarization around it [3]. In Bi2212, the $2\Delta/k_BT_c$ value for the SC mediated by spin-waves lies in strong coupling regime between 5.3 and 5.7 [2].

Let's consider different families of cuprates.

*Bi2212*. The Bi2212 has been studied using the tunneling spectroscopy [2]. The order parameter (OP) of the SC mediated by spin-waves has a d-wave symmetry. In Bi2212, the symmetry of the OP of the spinon SC has either a s-wave or mixed (s+d) symmetry [2, 6]. Figure

1 shows the two gaps on the Fermi surface in overdoped Bi2212 with $T_c$ = 89 K. The maximum magnitude of the tunneling gap corresponds to the spinon-gap. The maxima of the two gaps are shifted by 45°. In spite of the fact that the magnitude of the spinon-gap is larger than the magnitude of the MP-gap, the d-wave MP-gap is more intense than the spinon-gap and hence predominant in Bi2212 [2]. Figure 2 shows the phase diagram in Bi2212 obtained from the tunneling spectroscopy [2]. The hole concentration, $p$ has been calculated from the empirical relation $T_c/T_{c, max}$ = 1 - 82.6$(p - 0.16)^2$ which is satisfied for a number of HTSCs and we use $T_{c, max}$ = 95 K. The straight line corresponding to the spinon SC and the magnitude of the d-wave MP-gap in Bi2212 can be expressed by $\Delta$(meV) = 70.5 - 60$(p - 0.05)/0.22$ and $\Delta/\Delta_{max}$ = 1 - 82.6$(p - 0.16)^2$, respectively [2]. If we use the $2\Delta/k_BT_c$ value equal to 5.7 in Bi2212 then $\Delta_{max}$ = 23.3 meV. It is not clear yet if the spinon SC exists on stripes below $p$ = 0.05 and above $p$ = 0.27. It is possible that there is a collapse of the spinon SC at these points. One can see in Fig. 2 that the magnitude of the spinon-gap is always larger than the magnitude of the MP-gap. Since the coherence length $\xi \sim 1/\Delta$, hence, the coherence length of spinon pairs is always shorter approximately twice than the coherence length of MP pairs. In addition to main peaks in tunneling spectra, there is a sub-gap structure. At this moment, it is not clear yet what origin of this sub-gap. The d-wave SC mediated by spin-waves can coexist with a g-wave SC mediated by spin-waves [7, 8]. I have associated this sub-gap with the g-wave gap shown in Fig. 2. However, more research is needed to find out the exact origin of this sub-gap. In Bi2212, below $T_c$, there exists also the large normal-state spin-gap due to AF correlations [2]. Further, we will discuss only the spinon-gap and d-wave MP-gap.

The SC in cuprates occurs in $CuO_2$ planes which are basically identical in all cuprates. The maximum magnitude of the tunneling gap corresponds to the spinon-gap. It was shown that the maximum magnitude of tunneling gap in Bi2212 is not affected by the Pb-doping between the $CuO_2$ planes [9]. This means that the spinon SC is mainly defined by properties of $CuO_2$ planes. Since the $CuO_2$ planes are identical in all cuprates, it implies that the dependence of the magnitude of the spinon-gap on hole concentration will be in the first approximation the same in all cuprates including $YBa_2Cu_3O_{7-x}$ (YBCO) and NCCO (!).

*NCCO*. The OP in electron-doped NCCO has an anisotropic s-wave symmetry [10, 11]. This means that the SC mediated by spin-waves is absent in NCCO. Spin-waves for some reasons can not propagate into $CuO_2$ planes in NCCO (see discussion in Ref. 2). Nature had to look for another mechanism in order to establish the coherent-SC state in NCCO. I proposed earlier that the Josephson coupling between the stripes is responsible for it [2]. Since I realized that the spinon SC is basically the same in all cuprates including NCCO, the Josephson coupling between stripes can not be responsible for the coherent state in NCCO. The distance between stripes (~ 15 - 20 Å) is too large in comparison with the coherent length of spinon pairs ( 5 - 10 Å). The Josephson coupling energetically is most favorable way to establish the coherent state. Consequently, the SC mediated by spin-waves in hole-doped cuprates would not occur at all if the spinon SC could establish the coherent state due to the Josephson coupling. Because the spinon SC is the same in all cuprates, this means that the Josephson coupling between stripes can not be responsible for the coherent-SC state in NCCO neither. Another mechanism has to be involved in order to establish the coherent state for the spinon SC in NCCO. The interpretation of Raman scattering experiments [12] suggests that the SC in NCCO is mediated by phonons. Indeed, the in-plane normal-state resistivity in NCCO varies almost quadratically [13] and not linear with temperature like in other cuprates. Other important characteristics in NCCO are closer to classical SC behavior than in any other high-$T_c$ compound [12]. It seems that Nature chose phonons like in conventional SCs to be responsible for the coherent-SC state in NCCO. So, instead of the SC mediated by spin-waves, which occurs in hole-doped cuprates, the SC mediated by phonons dictates the value of $T_c$ in NCCO. This is the reason why the $T_c$ in NCCO is so low (22 K) and why the OP in NCCO has an anisotropic s-wave symmetry. This means that the phase diagram for NCCO will look approximately the same (with the exception of the g-wave gap which will be absent) as for Bi2212 shown in Fig. 2. Further, we will call the SC mediated by phonons as the conventional SC. The magnitude of the conventional-SC gap can be probably expressed also by $\Delta/\Delta_{max} = 1 - 82.6(p - 0.16)^2$. The main difference with Bi2212 will be in $\Delta_{max}$ which is equal to 4.6 - 5 meV in NCCO [10, 12].

*YBCO*. The YBCO is the only cuprate which has Cu-O chains. There is an evidence of the presence of the OP with a d-wave symmetry [11] like in Bi2212. Because of chains the phase diagram for YBCO will be different in a sense that in addition to the spinon-SC and d-g-wave MP-SC curves it will have one more curve corresponding to the SC on chains. The SC on chains occurs due to pairing of spinons, however, they can not acquire a gap above $T_c$ since the large normal-state spin-gap is a part of $CuO_2$ planes. The chains lie between planes. The "induced" SC state on chains occurs at $T$ $T_c$. This implies that the magnitude of the spinon-gap on chains will be equal to zero at $p = 0.05$ and 0.27. Consequently, the shape of the curve corresponding to the magnitude of the spinon-gap on chains will be similar to the d-wave MP-gap.

*LSCO*. In $La_{2-x}Sr_xCuO_4$ (LSCO), there is an evidence of the presence of the OP with a d-wave symmetry [11] like in YBCO and Bi2212. On the other hand, phonons couple with electrons in the SC state and thus participate in the electron-pairing interactions in LSCO [14]. Thus, in LSCO, the coherent state of the spinon SC is established via two channels which exist in parallel: a predominant SC mediated by spin-waves and much weaker SC mediated by phonons. This means that the spinon SC and SCs mediated by spin-waves and phonons coexist into $CuO_2$ planes in LSCO and hence in all cuprates since the $CuO_2$ planes are basically identical in all cuprates. The presence of the spinon SC on stripes is absolutely necessary. The coherent state the spinon SC into $CuO_2$ planes in all cuprates is established due to the SC mediated by spin-waves or/and phonons. In LSCO, the two channels coexist (compete) with each other. In NCCO, the magnetic channel is completely absent.

*OP.* The OP is not universal in all cuprates! From the discussion above, it is clear that the OP has to be different for each family of cuprates varying from a pure s-wave in NCCO to a predominant d-wave in Bi2212 and YBCO. In cuprates with high value of $T_c$ the OP will have a predominant d-wave of the magnetic origin because the high value of $T_c$ is the consequence of the presence of the SC mediated by spin-waves, which has the d-wave OP. The OP of the spinon SC has most likely an extended s-wave symmetry or varies from a s-wave to mixed (s+d) symmetry in different families of cuprates.

Into $CuO_2$ planes, there exist three SCs: the spinon SC, the SC mediated by spin-waves and SC mediated by phonons. Who would imagine earlier that three different SC mechanisms coexist in cuprates. Such diversity of different SC mechanisms is a consequence of charge inhomogeneity in cuprates.

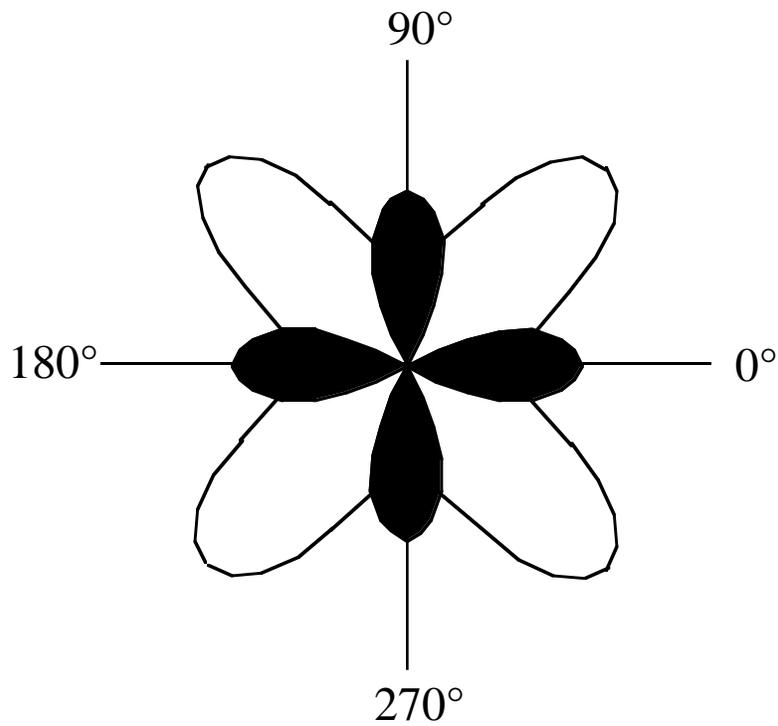

FIG. 1. Shapes of two superconducting gaps on the Fermi surface in Bi2212: black area (d-wave gap of magnetic polarons) and outlined area (spinon-gap having either s-wave or mixed (s+d) symmetry). The shapes of two gaps are shown schematically, however, the ratio between maximum magnitudes of two gaps corresponds to the real case with the hole concentration $p$ = 0.19 ( $T_c$ = 89 K) [2].

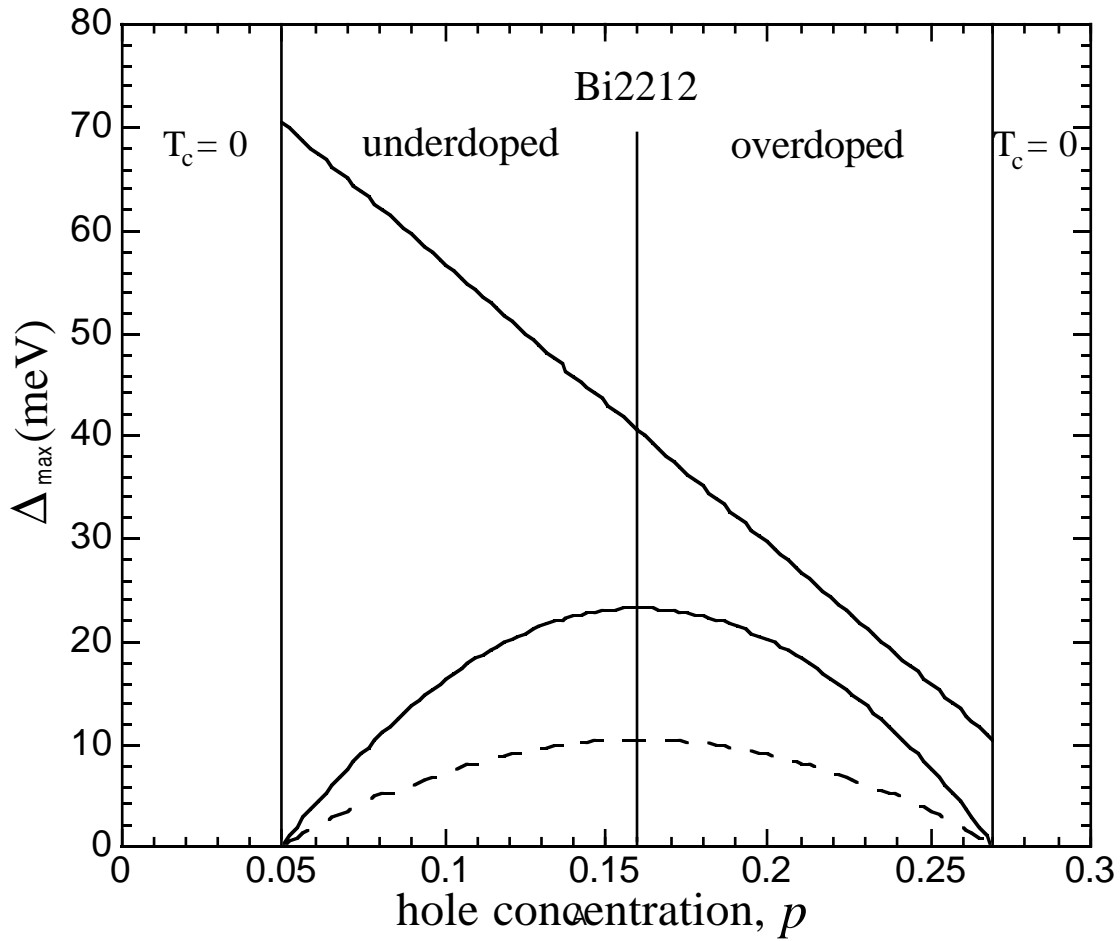

FIG. 2. Maximum SC gap values at 15 K vs. hole concentration in Bi2212 single crystals: straight line (spinon-gap on stripes, at $(\pi/2, \pi/2)$ on the Fermi surface); parabolic line (d-wave magnetic-polaron-gap at $(0, \pi)$), and dashed line (g-wave magnetic-polaron-gap at $(0, \pi)$) [2].